\documentstyle{a4w}

\newcommand{\nl}{{\hfill \break} }
\newcommand{\np}{ {\newpage } }

\newcommand{\N}{ \mbox{\rm I$\!$N} }
\newcommand{\R}{ \mbox{\rm I$\!$R} }

\newcommand{\lcm}{ \mbox{\rm lcm} }

\newcommand{\sign}{ \mbox{\rm sign} }

\newcommand{\artanh}{ \mbox{\rm artanh} }
\newcommand{\arcoth}{ \mbox{\rm arcoth} }

\newcommand{\MS}{ {\cal M\!S} }
\newcommand{\aBlLP}{1}
\newcommand{\bBlLP}{2}
\newcommand{\Pa}{3}
\newcommand{\Sch}{{4}}
\newcommand{\Li }{{5}}
\newcommand{\aBlZ }{6}
\newcommand{\Ma }{{7}}
\newcommand{\Xa }{{8}}
\newcommand{\Ga }{{9}}
\newcommand{\BlRZ }{{10}}
\newcommand{\Gi }{{11}}
\newcommand{\Vi }{{12}}
\newcommand{\aRa}{{13}}
\newcommand{\Ze }{{14}}
\newcommand{\bRa}{{15}}
\newcommand{\bBlZ}{{16}}

\begin{document}
\vspace*{-2.0cm}
\hfill {\bf July 1994}

\bigskip

\bigskip

\centerline{\Large \bf
On Singularities and Instability for Different Couplings}
\bigskip
\centerline{\Large \bf
between Scalar Field and Multidimensional Geometry\footnote
{This paper is dedicated to Prof. D. D. Ivanenko on the occasion of
his 90th birthday.}
}

\bigskip
\bigskip

\centerline{\bf U. Bleyer, M. Rainer}

\centerline{
Gravitationsprojekt, Universit\"at Potsdam}
\centerline{An der Sternwarte 16}
\centerline{D-14482 Potsdam, Germany}


\begin{abstract}

\noindent
We consider a multidimensional model of the universe given as a
$D$-dimensional geometry, represented by a Riemannian manifold $(M,g)$
with arbitrary signature of $g$, $M= \R\times M_1\times \cdots \times
M_n$, where the $M_i$ of dimension $d_i$ are Einstein spaces, compact
for $i>1$. For Lagrangian models $L(R,\phi)$ on $M$ which depend only
on the Ricci curvature $R$ and a scalar field $\phi$, there exists a
conformal equivalence with minimal coupling models. For certain
nonminimal models we study classical solutions and their relation to
solutions in the equivalent minimal coupling model. The domains of
equivalence are separated by certain critical values of the scalar
field $\phi$. Furthermore, the coupling constant $\xi$ of the coupling
between $\phi$ and $R$ is critical at both, the minimal value $\xi=0$
and the conformal value $\xi_c=\frac{D-2}{4(D-1)}$. In different
noncritical regions of $\xi$ the solutions behave qualitatively
different. Instability can occure only in certain ranges of $\xi$.
\end{abstract}

\section{Introduction}
\setcounter{equation}{0}
Gravitational models of multidimensional universes receive increasing
interest, since they provide a class of minisuperspace models, which
is rich enough to study the relation and the imprint of internal
compactified extra dimensions$^{\aBlLP,\bBlLP}$ on the external
space-time. In this paper we will consider classical multidimensional
models with respect to there properties in dependence on the form of
the coupling between geometry and a scalar field. The $D$-dimensional
geometry is represented by a Riemannian manifold $(M,g)$ with either
Lorentzian or Euclidean signature of $g$, $M= \R\times M_1\times
\cdots \times M_n$ with $D=1+\sum^n_{i=1} d_i$, where the $M_i$ of
dimension $d_i$ are Einstein spaces, compact for $i>1$. The first of
these spaces is to be considered as our observable exterior space,
while all the other factors represent internal spaces to be hidden at
present time.

In Sec. 2 we compare conformal transformations of Lagrangian models to
conformal coordinate transformations for a $D$-dimensional
geometry and define conformal equivalence.

In Sec. 3 we examine relations between conformally equivalent
Lagrangian models for $D$-dimensional geometry coupled to a spacially
homogeneous scalar field. Here, the conformal coupling constant
$\xi_c$ plays a distinguished role. We consider as example of special
interest the conformal transformation between a model with minimally
coupled scalar field  (MCM) and a conformally equivalent model with a
conformally coupled scalar field (CCM), thus generalizing previous
results from Refs. \Pa\ and \Sch, obtained for $n=1$ and $D=4$.

Sec. 4 introduces multidimensional cosmological models.

In Sec. 5 natural time gauges for multidimensional universes given by
the choices of i) the synchronous time $t_s$ of the universe $M$, ii)
the conformal time $\eta_i$ of a universe with the only spacial factor
$M_i$, iii) the mean conformal time $\eta$, given differentially as
some scale factor weighted average of $\eta_i$ over all $i$ and iv)
the harmonic time $t_h$, which will be used as specially convenient in
calculations on minisuperspace, since in this gauge the minisuperspace
lapse function is $N\equiv 1$.

In Sec. 6 the considerations on conformally equivalent Lagrangian
models from Sec. 3 will be pursued for the analysis of
multidimensional cosmological models on the level of solutions.

Sec. 7 is devoted to the distinguished role of the conformal coupling
constant $\xi_c=\frac{D-2}{4(D-1)}$ as compared to other values of $\xi$,
with special emphasis to stability considerations. Number theoretical
analysis of $\xi_c$, gives some hint, that it might play a crucial
role in the dimensional reduction process, since $\xi_c$ is especially
simple for $D=3,4,6,10$ and $26$. This is important for the cosmic
evolution, since it was e.g. shown in Ref. $\Li$ that inflation depends
critically on the number of dimensions.

Sec. 8 approaches the question of stability in dependence of the
dimension $D$ for a simple model$^\aBlZ$ with static internal spaces
instability conditions for the ground state of a scalar field. The
imprint of the extra dimensions yields conditions, which explictly
contain the conformal coupling constant $\xi_c$.

Sec. 9 resumes the perspective of the present results.

\section{ Conformal Transformations}
\setcounter{equation}{0}
One remark ab initio: throughout the following, on a
geometry $g$ on $M$, conformal transformations will actually
be represented as local Weyl transformations $g\mapsto e^{2f}g$ with
$f\in C^\infty (M)$.

We consider a differentiable manifold $M$. Equipped with a Riemannian
structure $g_{ij}$ and scalar fields $(\phi^1,\ldots,\phi^k)$ on $M$
we obtain a {\em Lagrangian model} by imposing a Lagrangian variation
principle
\begin{equation}
\delta S=0 \quad{\rm with}\quad S=\int_{M}  \sqrt{\vert g\vert}{L} d^D\!x
\end{equation}
given by a second order Lagrangian
\begin{equation}
L=L(g_{ij},\phi^1,\ldots,\phi^k;
    g_{ij,l},\phi^1_{,l},\ldots,\phi^k_{,l};
    g_{ij,lm}).
\end{equation}
Generally, we have to distinguish between  conformal
{\em coordinate} transformations in $D$-dimensional geometry and
conformal transformations of {\em Lagrangian models} for
$D$-dimensional geometry.\\

\noindent
{\bf Conformal transformation to new coordinates:}
\nl
We {\em fix a Lagrangian model} and transform the metric tensor components
conformally,
\begin{equation}
g_{i'j'}=e^{2f(x)}g_{kl},
\end{equation}
via a coordinate transform satisfying
\begin{equation}
dx'^{i}=e^{-f(x)}dx^i \quad{\rm or}\quad
\frac{\partial x'^{i}}{\partial x^j}=e^{-f(x)}\delta^i_j.
\end{equation}
Due to general covariance the model is still the same, though looking
different in  different coordinate frames. \\

\noindent
{\bf Conformal transformations of Lagrangian models:}
\nl
Conformal transformation {\em of the Lagrange model} keeps $M$ fixed as a
differentiable manifold,
but varies its additional structures conformally
\begin{equation}
(g_{ij},\phi^1,\ldots,\phi^k)\to (\hat g_{ij},\hat\phi^1,\ldots,\hat\phi^k),
\end{equation}
yielding a new variational principle by demanding
\begin{equation}
\sqrt{\vert g\vert}{L} \stackrel{!}{=} \sqrt{\vert \hat g\vert}{\hat L}
\end{equation}
for the new Lagrangian
\begin{equation}
\hat L=\hat L(\hat g_{ij},\hat\phi^1,\ldots,\hat\phi^k;
\hat g_{ij,l},\hat\phi^1_{,l},\ldots,\hat\phi^k_{,l};
\hat g_{ij,lm}).
\end{equation}
The action remains the same but the Lagrange density
becomes a new functional. Therefore, conformal transformations of models are
performed in practice on a {\em fixed coordinate patch} $x^i$ of $M$.
Lagrange models related in this way by a conformal transformation are
called conformally equivalent.

\section{ Conformally Equivalent Lagrangian Models}
\setcounter{equation}{0}

In this section we study transformations from a Lagrangian model with
minimally coupled scalar field (MCM) to a conformally equivalent one
with nonminimal coupled scalar field and vice versa.

Let us follow Ref. \Ma\  and consider an action of the kind
\begin{equation}
S=\int d^Dx\sqrt{\vert
g\vert}(F(\phi,R)-\frac{\epsilon}{2}(\nabla\phi)^2),
\end{equation}
where $F(\phi,R)$ in general describes nonminimal coupling.
With
\begin{equation}
\omega:=\frac{1}{D-2}\ln(2\kappa^2
                          \vert\frac{\partial F}{\partial R}\vert)+C
\end{equation}
the conformal factor
\begin{equation}
e^{\omega}=
[2\kappa^2\vert\frac{\partial F}{\partial R}\vert]^\frac{1}{D-2}e^{C}
\end{equation}
yields a conformal transformation from $g_{\mu\nu}$
to the MCM metric
\begin{equation}
\hat g_{\mu\nu}=e^{2\omega}g_{\mu\nu}.
\end{equation}
We note that CCM quantities $x$ correspond to MCM quantities $\hat x$
and the scalar field $\phi$ to $\Phi$ respectively.

Especially, let us consider actions being linear
in $R$:
\begin{equation}
S=\int d^Dx\sqrt{\vert g\vert}(f(\phi)R-V(\phi)
-\frac{\epsilon}{2}(\nabla\phi)^2).
\end{equation}
The MCM metric is then related to the metric of the conformal coupling
model (CCM) by
(3.4)
with
\begin{equation}
\omega=\frac{1}{D-2}\ln(2\kappa^2
\vert f(\phi)\vert)+C
\end{equation}
The scalar field in the MCM is
$$
\Phi=\kappa^{-1}\int d\phi\{\frac{\epsilon(D-2)f(\phi)+2(D-1)(f'(\phi))^2}
                               {2(D-2)f^2(\phi)}\}^{1/2}  =
$$
\begin{equation}
=(2\kappa)^{-1}\int d\phi\{\frac{2\epsilon f(\phi)+\xi_c^{-1}(f'(\phi))^2}
                               {f^2(\phi)}\}^{1/2},
\end{equation}
where
\begin{equation}
\xi_c:=\frac{D-2}{4(D-1)}
\end{equation}
is the conformal coupling constant.

For the following we define $\sign x$ to be $\pm 1$ for $x\geq 0$ resp.
$x<0$.
Then, with the new minimally coupled potential
\begin{equation}
U(\Phi)=(\sign f(\phi))\ [2\kappa^2\vert f(\phi)\vert]^{-D/D-2}V(\phi)
\end{equation}
the corresponding minimal action is
\begin{equation}
S=\sign f\int d^Dx\sqrt{\vert \hat{g}\vert}\left(-\frac{1}{2}
          [(\hat{\nabla}\Phi)^2-\frac{1}{\kappa^2}\hat{R}]-U(\Phi)\right).
\end{equation}
To be explicit, we will concentrate now on a model with
\begin{equation}
f(\phi)=\frac{1}{2}(1-\xi\phi^2),
\end{equation}
\begin{equation}
V(\phi)=\Lambda.
\end{equation}
Then the constant potential $V$ has its minimal correspondence in a
non constant $U$, given by
\begin{equation}
U(\Phi)=\pm \Lambda \vert\kappa^2 (1-\xi\phi^2) \vert^{-D/D-2}
\end{equation}
respectively
for $\phi^2<\xi^{-1}$ or $\phi^2>\xi^{-1}$.
Let us set in the following $\epsilon=1$. Then with
\begin{equation}
f'(\phi)=-\xi\phi
\end{equation}
we obtain
\begin{equation}
\Phi=\kappa^{-1}\int d\phi\{\frac{1+c\,\xi\phi^2}
                               {(1-\xi\phi^2)^2}\}^{1/2},
\end{equation}
where
\begin{equation}
c:=\frac{\xi}{\xi_c}-1.
\end{equation}
For $\xi=0$ it is $\Phi=\kappa^{-1}\phi +A$, i.e. the coupling remains
minimal.
To solve this integral for $\xi\neq 0$, we substitute $u:=\xi\phi^2$.

To assure a solution of
(3.15)
to be real, let us assume $\xi\geq\xi_c$
yielding $c\geq 0$. Then we obtain
$$
\Phi=\frac{\sign(\phi)}{2\kappa\sqrt{\xi}}
                  \int{\frac {\sqrt {u^{-1}+c}}{\vert1-u\vert}}du
+C_{<\atop>}
$$
$$
=\frac{\sign((1-\xi\phi^2)\phi)}{2\kappa\sqrt{\xi}}
\ln
\frac{ [2\,\sqrt {1+c} \sqrt{1+c\xi\,\phi^{2}}\sqrt {\xi}\vert\phi\vert
            +(2\,c+1)\xi\,\phi^{2}+1]^{\sqrt {1+c}}  }
{ [2\,\sqrt {c}\sqrt {1+c\xi\,\phi^{2}}\sqrt {\xi}\vert\phi\vert
             +2\,c\xi\,\phi^{2}+1]^{\sqrt{c}}
\cdot {\vert 1-\xi\,\phi^{2}\vert}^{\sqrt {1+c}}  }
$$
\begin{equation}
+C_{<\atop>}.
\end{equation}
The integration constants $C_{<\atop>}$
for $\phi^2<\xi^{-1}$ and $\phi^2>\xi^{-1}$ respectively
may be arbitrary functions of $\xi$ and the dimension $D$.
The singularities of the transform $\phi\to\Phi$ are located at
$\phi^2=\xi^{-1}$.

If the coupling is conformal $\xi=\xi_c$,
i.e. $c=0$, the expressions
(3.17)
simplify to
\begin{equation}
\kappa\Phi=\frac{1}{\sqrt{\xi_c}} [(\artanh\sqrt{\xi_c}\phi)+c_<]
\end{equation}
for $\phi^2<\xi^{-1}_c$ and to
\begin{equation}
\kappa\Phi=\frac{1}{\sqrt{\xi_c}} [(\arcoth\sqrt{\xi_c}\phi)+c_>]
\end{equation}
for $\phi^2>\xi^{-1}_c$.
Then the inverse formulas expressing the conformal field $\phi$ in terms of
the minimal field $\Phi$ are
\begin{equation}
\phi=\frac{1}{\sqrt{\xi_c}} \left[ \tanh(\sqrt{\xi_c}\kappa\Phi-c_<) \right]
\end{equation}
with $\phi^2<\xi^{-1}_c$ and
\begin{equation}
\phi=\frac{1}{\sqrt{\xi_c}} \left[ (\coth(\sqrt{\xi_c}\kappa\Phi-c_>) \right]
\end{equation}
with $\phi^2>\xi^{-1}_c$ respectively.
This result agrees with Ref. \Xa.
For $D=4$ it has been obtained earlier in Refs. \Pa, \Sch\ and \Ga.
In Ref. \Ga\ it has been shown for $D=4$,  that while the MCM
shows a curvature singularity, the conformal coupling
model with $\phi$ of Eq.
(3.20)
has no such singularity.

The conformal factor is according to Eqs.
(3.6)
and
(3.11)
given by
\begin{equation}
\omega=\frac{1}{D-2}\ln(\kappa^2 \vert 1-\xi_c\phi^2 \vert)+C.
\end{equation}

The singularity of the conformal transformation
(3.22)
at $\phi^2=\xi^{-1}_c$ separates different regions in $\phi$ where
conformal equivalence between the MCM and CCM holds.
Eqs.
(3.20)
and
(3.21)
illustrate the qualitatively
different behavior in the two regions.
In Ref. \BlRZ\  qualitative differences have also been found
in multidimensional solutions of the respective models.

\section{ Lagrangian Models for Multidimensional Cosmology}
\setcounter{equation}{0}
We consider a geometry described by a (Pseudo-) Riemannian manifold
$$
M=\R\times M_1 \times\ldots\times M_n,
$$
with the first fundamental form
\begin{equation}
g\equiv ds^2 = -e^{2\gamma} dt\otimes dt
     + \sum_{i=1}^{n} a_i^2 \, ds_i^2,
\end{equation}
where    $ a_i=e^{\beta^i} $
is the scale factor of the $d_i$-dimensional space $M_i$
with the first fundamental form
\begin{equation}
ds_i^2
=g^{(i)}_{{k}{l}}\,dx^{k}_{(i)} \otimes dx^{l}_{(i)}.
\end{equation}
If we assume within a multidimensional geometry
(4.1)
that the $M_i$ are Einstein spaces of constant curvature,
then the Ricci scalar curvature of $M$ is
\begin{equation}
R=e^{-2\gamma}\left \{
\left [ \sum_{i=1}^{n} (d_i \dot\beta^i) \right ]^2
+ \sum_{i=1}^{n}
d_i[ {(\dot\beta^i)^2-2\dot\gamma\dot\beta^i+2\ddot \beta^i} ]
\right \}
+\sum_{i=1}^{n} R^{(i)} e^{-2\beta^i}.
\end{equation}

Let us now consider a variation principle with the action
\begin{equation}
S=S_{EH}+S_{GH}+S_{M},
\end{equation}
where
\[
S_{EH}=\frac{1}{2\kappa^2}\int_{M}\sqrt{\vert g\vert} R\, dx
\]
is the Einstein-Hilbert action, $S_{GH}$
is the Gibbons-Hawking boundary term$^\Gi$,
 and $S_M$ the action of matter.

Let us consider the matter given by  a minimally coupled
scalar field $\Phi$ with potential $U(\Phi)$.
Then the
variational principle of
(4.4)
is  equivalent to
a Lagrangian variational principle
over the minisuperspace $\cal M$, which is spanned by the $\beta^i$,
and the scalar field $\Phi$,
$$
S=\int Ldt,
$$
\begin{equation}
L=\frac{1}{2}{\mu}
e^{-\gamma+\sum_{i=1}^{n}d_i\beta^i}
\left \{
\sum_{i=1}^{n}{d_i(\dot\beta^i)^2}
-[\sum_{i=1}^{n}{d_i\dot\beta^i}]^2
+\kappa^2{\dot \Phi}^2
\right \} - V(\beta^i, \Phi)
\end{equation}
with
$$
V(\beta^i, \Phi) = - \frac{1}{2}{\mu}
e^{\gamma+\sum_{i=1}^{n}d_i\beta^i}R^{(i)}e^{-2\beta^1}
+ {\mu}\kappa^2 e^{\gamma+\sum_{i=1}^{n}d_i\beta^i}
U(\Phi),
$$
where
\begin{equation}
\mu:=\kappa^{-2}\prod_{i=1}^{n}\sqrt{\vert\det g^{(i)}\vert}.
\end{equation}
It is a convenient proceedure of cosmologists,
to extend the minisuperspace $\cal M$ of pure geometry directly by an
additional dimension from the scalar field $\Phi$ as further
minisuperspace coordinate, yielding an enlarged
minisuperspace $\MS$.

Let us define a metric on $\MS$, given in
coordinates $\beta^i$, $i=1,\ldots,n+1$ with $\beta^{n+1}:=\kappa\Phi$.
We set
\begin{equation}
G_{n+1\,i}=G_{i\,n+1}:=\delta_{i\,n+1}
\ \mbox{and}\
G_{kl}:=d_k \delta_{kl}-d_k d_l
\end{equation}
for $i=1,\ldots,n+1$ and $k,l=1,\ldots,n$,
thus defining the components
$G_{ij}$ of the minisuperspace metric
\begin{equation}
G = G_{ij}d\beta^i\otimes d\beta^j.
\end{equation}
Then we get the Lagrangian
\begin{equation}
L=\frac {\mu}{2} G_{ij}\dot \beta ^i\dot \beta ^j
-V(\beta^i).
\end{equation}
with the energy constraint
\begin{equation}
\frac {\mu}{2} G_{ij}\dot \beta ^i\dot \beta ^j
+ V(\beta^i) = 0.
\end{equation}
Independent global conformal tranformations of
the spaces $M^{(i)}$ yield just translations in the
functions $\beta^i$.

\section{ Natural Times in Multidimensional Geometry}
\setcounter{equation}{0}
For this geometry let us
compare different choices of time $\tau$  in Eq.
(4.1).
The time gauge is determined by the function $\gamma$.
There exist few  time gauges, natural from the physical point of view.

i) The {\em synchronous time gauge}
\begin{equation}
\gamma\equiv 0,
\end{equation}
for which $t$ in Eq.
(4.1)
is the proper time $t_s$ of the universe.
The clocks of geodesically comoved observers go synchronous to that
time.

ii) The {\em conformal time gauges} on $\R\times M_i\subset M$
\begin{equation}
\gamma\equiv \beta^i,
\end{equation}
for which $t$ in Eq.
(4.1)
is the conformal time $\eta_i$ of $M_i$
for some $i\in \{1,\ldots,n\}$,
given by
\begin{equation}
d\eta_i=e^{-\beta^i}dt_s.
\end{equation}

iii) The {\em mean conformal time gauge} on $M$:
\nl
For $n>1$ and $\beta^2\neq\beta^1$
on $M$ the usual concept of a conformal time does no longer apply.
Looking for a generalized ``conformal time" $\eta$ on $M$, we set
\begin{equation}
d:=D-1=\sum_{i=1}^n d_i
\end{equation}
and consider the gauge
\begin{equation}
\gamma\equiv \frac{1}{d}\sum_{i=1}^n d_i\beta^i,
\end{equation}
which yields a time $t\equiv\eta$ given by
\begin{equation}
d\eta=\left( \prod_{i=1}^n a_i^{d_i} \right)^{-1/d} dt_s.
\end{equation}
Here $\prod_{i=1}^n a_i^{d_i}$ is proportional to the volume of
$d$-dimensional spacial sections in $M$
and the relative time scale factor
\begin{equation}
\left( \prod_{i=1}^n a_i^{d_i} \right)^{1/d}
=e^{\frac{1}{d}\sum_{i}d_i\beta^i}
\end{equation}
is given by a scale exponent, which is the dimensionally weighted arithmetic
mean of the spacial scaling exponents of spaces $M_i$.
It is
\begin{equation}
(dt_s)^d= e^{\sum_{i}d_i\beta^i} d\eta^d.
\end{equation}
On the other hand by Eq.
(5.3)
we have
\begin{equation}
(dt_s)^d=\otimes_{i=1}^n \left( e^{\beta^i}d\eta_i \right)^{d_i},
\end{equation}
and together with Eq.
(5.8)
we get
\begin{equation}
(d\eta)^d=e^{-\sum_{i}d_i\beta^i}
\otimes_{i=1}^n \left( e^{\beta^i}d\eta_i \right)^{d_i}.
\end{equation}
So the time $\eta$ is a mean conformal time, given differentially
as a dimensionally scale factor weighted geometrical tensor
average of the conformal times $\eta_i$.
An alternative to the mean conformal time $\eta$
is given by a similar differential averaging like Eq.
(5.10),
but weighted by an additional factor of $e^{(1-d)\sum_{i}d_i\beta^i}$.
This gauge is described in the following.

iv) The {\em harmonic time gauge}
\begin{equation}
\gamma\equiv\gamma_h:=\sum_{i=1}^n d_i\beta^i
\end{equation}
yields the time $t\equiv t_h$, given by
\begin{equation}
dt_h=\left( \prod_{i=1}^n a_i^{d_i} \right)^{-1} dt_s
=\left( \prod_{i=1}^n a_i^{d_i} \right)^{\frac{1-d}{d}} d\eta.
\end{equation}
In this gauge any function $\varphi$ with $\varphi (t,y)=t$
is harmonic, i.e. $\Delta[g] \varphi =0$, and
the minisuperspace lapse function is $N\equiv 1$.
This gauge is especially convenient when we work in
minisuperspace.

In application to the model of Sec. 4
note that, if e.g. time is harmonic in the MCM
\begin{equation}
\tau\equiv t^{(m)}_h,
\end{equation}
in the CCM it cannot be expected to be harmonic either,
i.e. in general
\begin{equation}
\tau\neq t^{(c)}_h.
\end{equation}
In general natural time gauges
are not preserved by conformal transformations
of the geometry.
They have to be calculated by a coordinate transformation
in each of the conformally equivalent models separately.

\section{Different Couplings in Multidimensional Cosmologies}
\setcounter{equation}{0}
In the following we want to pursue the comparison
of the MCM and the CCM
on the level of their classes of solutions for a
multidimensional geometrical model of cosmology.
Let us specify the geometry for the MCM to be of multidimensional
type
(4.1),
with all $M_i$, $i=1,\ldots,n$, being Ricci flat
and the minimally coupled scalar field to have zero potential
$U\equiv 0$.
In the harmonic time gauge
(5.11)
with harmonic time
\begin{equation}
\tau\equiv t^{(m)}_h,
\end{equation}
we demand this model to be a solution for Eq.
(4.9)
with vanishing $R^{(1)}$ and $U(\Phi)$
with $\beta^{n+1} = \kappa \Phi$. This solution is
a multidimensional Kasner like universe,
given by
\begin{equation}
\hat\beta^i=b^i\tau+c^i  \ \mbox{and}\
\hat\gamma=\sum_{i=1}^n d_i \hat\beta^i
=(\sum_{i=1}^n d_i b^i)\tau+(\sum_{i=1}^n d_i c^i),
\end{equation}
with $i=1,\ldots,n+1$,
where with $V\equiv 0$ the constraint Eq.
(4.10)
simply reads
\begin{equation}
G_{ij} b^i b^j + (b^{n+1})^2=0.
\end{equation}
With Eq.
(3.22)
the scaling powers of the universe given by Eqs.
(6.2)
with $i=1,\ldots,n$ transform to corresponding
scale factors of the CCM universe
$$
\beta^i=\hat\beta^i-\omega
$$
\begin{equation}
=b^i\tau+\frac{1}{2-D}\ln\vert 1-\xi_c(\phi)^2\vert
+ c^i + \frac{2}{2-D} \ln\kappa-C
\end{equation}
and
$$
\gamma=\sum_{i=1}^n d_i \beta^i
$$
\begin{equation}
=(\sum_i d_i b^i)\tau+\frac{1}{2-D}\ln\vert 1-\xi_c(\phi)^2\vert
+(\sum_i d_i c^i) + \frac{2}{2-D} \ln\kappa-C.
\end{equation}
Let us take for simplicity
\begin{equation}
C=\frac{2}{2-D} \ln\kappa,
\end{equation}
which yields the lapse function
\begin{equation}
e^\gamma=e^{(\sum_i d_i b^i)\tau+(\sum_i d_i c^i)}
\vert 1-\xi_c(\phi)^2\vert^{\frac{1}{2-D}}
\end{equation}
and for $i=1,\ldots,n$ the scale factors
\begin{equation}
e^{\beta^i}= e^{b^i\tau+ c^i}
\vert 1-\xi_c(\phi)^2\vert^{\frac{1}{2-D}}.
\end{equation}
Let us further set for simplicity
\begin{equation}
c_{<}=c_{>}=\sqrt{\xi_c} c^{n+1}.
\end{equation}

The transformation of the scalar field from the
solution
(3.22)
of the MCM
\begin{equation}
\kappa\Phi(\tau)=b^{n+1}\tau+c^{n+1}
\end{equation}
to the scalar field of the conformal model by Eqs.
(3.20)
resp.
(3.21)
and substitution of the latter in Eqs.
(6.7)
and
(6.8)
yields a lapse function
\begin{equation}
e^\gamma=e^{(\sum_i d_i b^i)\tau+(\sum_i d_i c^i)}
\cosh^{\frac{2}{D-2}}( \sqrt{\xi_c} b^{n+1}\tau )
\end{equation}
resp.
\begin{equation}
e^\gamma=e^{(\sum_i d_i b^i)\tau+(\sum_i d_i c^i)}
\vert\sinh^{\frac{2}{D-2}}( \sqrt{\xi_c} b^{n+1}\tau )\vert
\end{equation}
and, with $i=1,\ldots,n$, nonsingular scale factors
\begin{equation}
e^{\beta^i}= e^{b^i\tau+ c^i}
\cosh^{\frac{2}{D-2}}( \sqrt{\xi_c} b^{n+1}\tau )
\end{equation}
resp. singular scale factors
\begin{equation}
e^{\beta^i}= e^{b^i\tau+ c^i}
\vert\sinh^{\frac{2}{D-2}}( \sqrt{\xi_c} b^{n+1}\tau )\vert
\end{equation}
of the CCM. The scale factor singularity
of the MCM for $\tau\to-\infty$ vanishes in the CCM
of Eqs.
(6.11)
and
(6.13)
for a scalar field $\phi$ bounded
according to (3.20).
For $D=4$ this result had already been indicated by Ref. \Ga.

On the other hand in the CCM
of Eqs.
(6.12)
and
(6.14),
with $\phi$ according to
(3.21),
though the scale factor singularity of the minimal model
for $\tau \to -\infty$ has also disappeared, instead there is another new
scale factor singularity at finite
(harmonic) time $\tau=0$.

Let us consider a special case of the nonsingular solution
with $\phi^2<\xi_c^{-1}$,
where we assume the internal spaces
to be static in the MCM, i.e.  $b^i=0$ for $i=2,\ldots,n$.
Then in the CCM, the internal spaces are no longer
static. Their scale factors
(6.13)
with $i>2$ have a minimum at $\tau=0$. Remind that for solution
(6.2)
all spaces $M_i$, internal and external, $i=1,\ldots,n$ have
been assumed as flat. {}From Eq.
(6.3)
with $G_{11}=d_1(1-d_1)$ we find
that the scalar field is given by
\begin{equation}
(b^{n+1})^2= d_1 (d_1-1) (b^1)^2.
\end{equation}
With real $b_1$ then also
\begin{equation}
b^{n+1}=\pm\sqrt{d_1 (d_1-1)}b^1
\end{equation}
is real and by Eq.
(6.13)
the scale $a_1$ of $M_1$ has a minimum at
\begin{equation}
\tau_0
=(\sqrt{\xi_c} b^{n+1})^{-1}\artanh\left(
\frac{(2-D)}{2\sqrt{\xi_c}}
\frac{b^1}{b^{n+1}} \right),
\end{equation}
with $\tau_0>0$ for $b^1<0$ and $\tau_0<0$ for $b^1>0$.

The points $\tau=\tau_0$ and $\tau=0$ are the turning points
in the minimum for the factor spaces $M_1$ and $M_2,\ldots,M_n$
respectively.
It is interesting to explain the creation of our Lorentzian universe
by a ''birth from nothing''$^\Vi$, i.e. quantum tunneling from an Euclidean
region.
Let us first consider the geometry of this tunneling as usual for the
external universe $\R\times M_1$.
So if we cut $M$
along the minimal hypersurface at $\tau_0$ in 2 pieces, one of them,
say $M'$, contains
the hypersurface $\tau=0$ where the internal spaces are minimal.
We set $M'':=M\setminus M'$ to be the remaining piece.
Then we can choose (eventually with time reversal $\tau\to-\tau$)
either $M'$ or $M''$ as a universe $\tilde M$ that is generated at $\tau_0$
with initial minimal scale $a_1(\tau_0)$.
In the usual quantum tunneling interpretation, at the
scale $a_1(\tau_0)$ with $\dot a_1(\tau_0)=0$
one glues smoothly a compact simply connected Euclidean space-time
region to the Lorentzian $\tilde M$, yielding a joint differentiable
manifold $\hat M$. Then the sum of classical paths in $\hat M$
passing the boundary $\partial\tilde M$ from the Euclidean to the
Lorentzian region can be interpreted as quantum tunneling from
''nothing''$^\Vi$ to $\tilde M$.

According to Ref. $\aRa$ this interpretation has a direct topological
correspondence in a projective blow up of a singularity of shape
$M_2\times \cdots \times M_n$ (the ''nothing'') to
$S^{d_1}(a_1(\tau_0))\times M_2\times \cdots \times M_n$, where
$S^{d_1}(a_1(\tau_0))$ denotes the $d_1$-dimensional sphere of radius
$a_1(\tau_0)$.

For $\tilde M=M'$ the internal spaces shrink for (harmonic) time
from $\tau_0$ towards $\tau=0$ and expand from $\tau=0$ onwards for ever,
but for $\tilde M=M''$ the internal spaces expand for (harmonic) time
from $\tau_0$ onwards for ever.
So the decomposition of $M$ in $M'$ and $M''$ is highly asymmetric
w.r.t. the internal spaces. For more realistic
models it might be especially useful to consider
the piece of $M'$ which lies between $\tau_0$ and $\tau=0$,
since it can describe a shrinking of internal spaces while the
external space is expanding.

Remarkably, the multidimensional geometries
with $\tau<\tau_0$ and $\tau>\tau_0$ are $\tau$-asymmetric to each
other. Taking one as contracting, the other as expanding
w.r.t. $M_1$, the two are distinguished
by a qualitatively different behavior of internal spaces $M_k$, $k\geq 2$.

The latter allows to choose the ''arrow of time''$^\Ze$ in a natural
manner determined by  intrinsic features of the solutions.
Note, if there is at least one
internal extra space, i.e. $n>1$, then the minisuperspace w.r.t.
scalefactors of geometry
has Lorentzian signature $(-,+,\ldots,+)$. After diagonalization of
(4.7)
by a minisuperspace coordinate transformation $\beta^i\to\alpha^i$
($i=1,\ldots,n$), there is just one new scale factor coordinate, say
$\alpha^1$, which corresponds to the negative eigenvalue of $G$, and
hence assumes the role played by time in usual quantum mechanics. (For
$n=1$ there are no internal spaces, but $G_{11}<0$ for $d_1>1$ still
provides a negative eigenvalue that is distinguished at least against
the additional positive eigenvalue from the scalar field.) This shows
that, at least after diagonalization, an ''external'' space is
distinguished against the internal spaces, because its scale factor
provides a natural ''time'' coordinate.

Upto now we have considered the smooth tunneling from
an Euclidean region to the external universe $\R\times M_1$,
where the external spaces have been considered as purely
passive spectators of the tunneling process.
As we have pointed out in contrast to models with only one (external) space
factor $M_1$, the additional internal spaces $M_2,\ldots,M_n$
yield an asymmetry of $M$ w.r.t. (harmonic) time $\tau$ for
$\tau_0\neq 0$, which is according to Eq.
(6.17)
the case exactly when
$D\neq 2$ and the external space is non static, i.e. $b_1\neq 0$.

In the following we want to obtain a quantum tunneling interpretation
for all of $M$, including the internal spaces. The picture
becomes more complicated, since the extremal hypersurfaces of
external space and internal spaces are located at different times
$\tau=\tau_0$ resp. $\tau=0$.

Let $M_1$ be the external space with $b_1>0$ and hence $\tau_0<0$.
Let us start with an Euclidean region of complex geometry
given by scale factors
\begin{equation}
a_k=e^{-ib^k\tau+\tilde c^k}
\vert\sin ( \sqrt{\xi_c} b^{n+1}\tau )\vert^{\frac{2}{D-2}}.
\end{equation}
Then we can perform an analytic continuation to the Lorentzian region
with $\tau\to i\tau+\pi/(2 \sqrt{\xi_c} b^{n+1})$, and we require
$c^k=\tilde c^k-i\pi b^k/(2 \sqrt{\xi_c} b^{n+1})$ to be the real
constant of the real geometry
(6.8).

The quantum creation (via tunneling)
of different factor spaces takes place at different values of $\tau$.

First the factor space $M_1$ comes into real existence and after an
time interval $\Delta\tau=\vert\tau_0\vert$ the internal factor spaces
$M_2, \ldots, M_n$ appear in the Lorentzian region. Since $\Delta\tau$
is arbitrarily large, there is in principle an alternative explanation
of the unobservable extra dimensions, independent from concepts of
compactification and shrinking to a fundamental length in symmetry
breaking. Here, they may have been up to now still in the Euclidean
region and hence unobservable. This view is also compatible with the
interpretation$^\aRa$ of the internal symmetries as complex
resolutions of simple singularities of Cartan series ADE.

Now let us perform a transition from Lorentzian time $\tau$ to Euclidean
time $i\tau$. Then with a simultaneous transition from $b^k$ to
$- i b^k$
for $k=1,\ldots,n$ the geometry remains real, since
$\hat\beta^k=b^k\tau+c^k$ is unchanged.
But the analogue of Eq.
(6.16)
for the Euclidean region then becomes
\begin{equation}
b^{n+1}=\mp i\sqrt{d_1 (d_1-1)}b^1.
\end{equation}
Hence the scalar field is purely imaginary.
This solution corresponds to a classical (instanton) wormhole.
The sizes of the wormhole throats in the factor spaces $M_2,\ldots,M_n$
coincide with the sizes of static spaces in the
minimal model, i.e. $\hat a_2(0),\ldots,\hat a_n(0)$ respectively.

With Eq.
(6.16)
replaced by
(6.19),
the Eq.
(6.17)
remains unchanged in the transition to the Euclidean region,
and the minimum of the scale $a_1$ (unchanged geometry !)
now corresponds to the throat of the wormhole.

If one wants to compare the synchronous time pictures of the
MCM and the CCM solutions, one has to calculate them for
both metrics. In the MCM we have
\begin{equation}
dt^{(m)}_s=e^{\hat\gamma}d\tau
=e^{(\sum_i d_i b^i)\tau+(\sum_i d_i c^i)}d\tau,
\end{equation}
which can be integrated to
\begin{equation}
t^{(m)}_s=(\sum_i d_i b^i)^{-1}
e^{\hat\gamma}+t_0.
\end{equation}

The latter can be inverted to
\begin{equation}
\tau=(\sum_i d_i b^i)^{-1}
\left\{[\ln(\sum_i d_i b^i)(t^{(m)}_s-t_0)]-(\sum_i d_i c^i)\right\}.
\end{equation}
Setting
\begin{equation}
B:=\sum_{i=1}^{n} d_i b^i \ \mbox{and} \ C:=\sum_{i=1}^{n} d_i c^i,
\end{equation}
this yields the scale factors
\begin{equation}
\hat{a_s}^i=(t^{(m)}_s-t_0)^{b^i/B} e^{\frac{b_i}{B}(\ln B-C)+c_i}
\end{equation}
and the scalar field
\begin{equation}
\kappa\Phi=\frac{b^{n+1}}{B} \{ [\ln B(t^{(m)}_s-t_0)] -C \}+c^{n+1}.
\end{equation}
Let us define for $i=1,\ldots,n+1$ the numbers
\begin{equation}
\alpha^i:=\frac{b^i}{B}.
\end{equation}
With
(6.23)
they satisfy
\begin{equation}
\sum_{i=1}^{n} d_i \alpha^i=1,
\end{equation}
and by Eq.
(6.3)
also
\begin{equation}
\alpha^{n+1}=\sqrt{1-\sum_{i=1}^{n} d_i (\alpha^i)^2}.
\end{equation}
Eqs.
(6.24)
shows, that the solution
(6.2)
is a generalized Kasner universe with exponents $\alpha^i$
satisfying generalized Kasner conditions
(6.27) and (6.28).

In the conformal model the synchronous time is given as
\begin{equation}
t^{(c)}_s=\int e^\gamma d\tau=
\int\cosh^{\frac{2}{D-2}}(\sqrt{\xi_c} b^{n+1}\tau)
e^{B\tau+C} d\tau
\end{equation}
resp.
\begin{equation}
t^{(c)}_s=\int e^\gamma d\tau=
\int \sinh^{\frac{2}{D-2}}(\sqrt{\xi_c} b^{n+1}\tau)
e^{B\tau+C} d\tau.
\end{equation}
Similarily one could also try to calculate other time gauges for
both metrics.

\section{ $\xi_c$ and Resonance of Coupling in Different
Dimensions}
\setcounter{equation}{0}
In this section we
examine the resonance of the coupling in dpendence of the dimension.
First we consider only conformal coupling constants
$\xi_c=\frac{D-2}{4(D-1)}$
and study their prime factorization.
Table 1 lists $\xi_c=:\frac{r}{q}$ with
only trivial greatest common divisor of $r$ and $q$, i.e. gcd$(r,q)=1$,
$p_{max}$ the maximal primefactor contained in either $r$ or $q$ and
the least common multiple lcm$(\xi):=$lcm$(r,q)$,
for dimensions $D=3\ldots 30$.
{\small
$$
\begin{array}{lccccccccccccccc}
D                &:& 3& 4& 5& 6& 7& 8& 9&10&11&12&13&14&15&16\\
\xi_c=\frac{r}{q}&:&\frac{1}{8}&\frac{1}{6}&\frac{3}{16}&\frac{1}{5}&
\frac{5}{24}&\frac{3}{14}&\frac{7}{32}&\frac{2}{9}&\frac{9}{40}&
\frac{5}{22}&\frac{11}{48}&\frac{3}{13}&\frac{13}{56}&\frac{7}{30}\\
p_{max}          &:&2&3&3&5&5&7&7&3&5&11&11&13&13&7\\
\lcm             &:&8&6&48&5&120&42&224&18&360&110&528&39&728&210\\
\end{array}
$$
$$
\begin{array}{cccccccccccccc}
17&18&19&20&21&22&23&24&25&26&27&28&29&30\\
\frac{15}{64}&\frac{4}{17}&\frac{17}{72}&\frac{9}{38}&
\frac{19}{80}&\frac{5}{21}&\frac{21}{5}&\frac{11}{46}&\frac{23}{96}&
\frac{6}{25}&\frac{25}{104}&\frac{13}{54}&\frac{27}{112}&\frac{7}{29}\\
5&17&17&19&19&7&11&23&23&5&13&13&7&29 \\
960&68&1224&342&1520&105&1848&506&2208&150&2600&702&3024&203
\end{array}
$$
\begin{center}
Table 1: $p_{max}$ and lcm of $\xi_c$ for $D=3,\ldots,30$.
\end{center}
}
\nl\nl
Can we give an interpretation of this numbers in terms of higher
stability of certain dimensions with higher `simplicity' of $\xi$ ?

To answer this question we assume that a system is described by
a Lagrangian
$$
      L(q,\dot q):=L_1(q,\dot q)+\xi_c L_2(q,\dot q),
$$
where $q$ denotes the configuration variables. The conjugate momenta are
$$
p:=\frac{\partial L}{\partial \dot q },
$$
so the Legendre transform is given by the linear differential operator
$$
O_{\dot q}:=\dot q \frac{\partial}{\partial\dot q}-1
$$
applied to L. Since the operator $O_{\dot q}$ is linear we obtain
$$
H=O_{\dot q}L=O_{\dot q}L_1+\xi_c O_{\dot q}L_2 =H_1+\xi_c H_2.
$$
Now suppose the configuration variables to be (massless) scalar fields.
Their energy has to be quantized. Thus for the ground state $\vert0>$
we yield (in natural units with $\hbar=1$)
$$
<0\vert H\vert0>=\omega_1+\xi_c \omega_2
$$
with frequencies $\omega_{1,2}=<0\vert H_{1,2}\vert 0>$ respectively.
The latter are the better in resonance
the smaller the $\lcm$ of $\xi_c$ and the `simpler' the fraction $\xi_c$.
If we consider this resonance as supporting the evolution leading
to our present state, we find different stability of different dimensions,
according to the `simplicity' of $\xi_c$ or the smallness of the
$\lcm(\xi_c)$. {}From the table above we find $D=6$ most stable, followed by
$D=4$ and $D=3$. In these dimensions $\xi^{-1}_c$ is just an integer.
Besides $D=6,4,3$ the next best is $D=10$ with $\xi_c=\frac{2}{9}$.
Note that all dimensions of the form $D=4i+2, i\in\N$ are more stable than
their neighbouring dimensions. If we admit for the rational
composition only the first $3$ prime numbers, then in the range
$10<D<81$ the next best choice is $D=26$. While for $D=82=4\cdot 20+2$
also $p_max=5$, the corresponding $\lcm(\frac{20}{81})=1620$ is
already more than $10$ times higher than that of $D=26$.

For general coupling constants $\xi$ this considerations become relevant
for regions in which $\xi\to\xi_c$ assympotically. There might be dynamical
necessity to avoid this regions for their conformal resonances, and this
even more for the stable dimensions. Even more the stability of the
dimensions with high resonance might be due to an evolutionary effect
the avoiding of the resonances has on the dynamics.
For a better understanding of the latter is necessary to understand the
conditions for symmetry breaking.
A necessary condition for this is the instability of the vacuum.

\section{The Negative Mass Condition for Vacuum Instability}
\setcounter{equation}{0}

In the following we look for dependence of the vacuum instability on
the (not necessarily conformal) coupling constant in different dimensions
of external space $M_1$ in the case that $M_i$ for $i>1$ are all flat.

More specifically we consider a model where the internal spaces
$M_i$ for $i>1$ are Ricci flat and static.
The dynamics then is equivalent to that of a model on $\R\times M_1$.
We consider a scalar field on the background of the curved space-time $M$
or equivalently $\R\times M_1$. Note that in contrast to previous sections
here the backreaction of the scalar field on the geometry will be
assumed to be neglegible.
In the ground state through the evolution of $a_1$ the extra dimensions
leave a dynamical imprint in $M_1$.

A necessary condition for vacuum instability is
a negative effective mass
\begin{equation}
M^2<0.
\end{equation}

For the rest of this section we choose the conformal time gauge
\begin{equation}
\gamma = \beta^1,
\end{equation}
i.e. $t\equiv\eta_1$, the conformal time of the external space $M_1$.
Correspondingly, for any $x$ we define
\begin{equation}
\dot x:=\frac{\partial x}{\partial \eta_1},
\end{equation}
i.e. differently from the convention  of the previous sections
(where the dot denotes the
derivative w.r.t. the harmonic time), now and in the following
the dot denotes the partial derivative w.r.t. the conformal time of $M_1$.

The effective mass $M$ of the scalar field is given as$^\aBlZ$ ($a\equiv a_1$)
\begin{equation}
M^2=m^2a^2+\xi d(d-1)k
+d(\xi-\xi_c)(2\frac{\ddot a}{a}-(d-3)\frac{\dot a^2}{a^2}),
\end{equation}
where $m$ is the  bare mass.
In the following we assume for simplicity vanishing bare mass,
$m=0$, and
examine the condition
(8.4) in different cases depending on the
curvature $k$ of the  exterior space and the ratio $\frac{\ddot a}{a}$.
Let us substitute
\begin{equation}
A_d:=d(d-3)\frac{\dot a^2}{a^2}
\end{equation}
in Eq.
(8.3). Note that $A_d=0$ for critical dimension $d=3$ and
$A_d {>\atop<} 0$ for $d {>\atop<} 3$ respectively.

With Eq.
(8.5) we obtain
$$
M^2=k d(d-3)\xi
+d(\xi-\xi_c)(2\frac{\ddot a}{a}-(d-3)(\frac{\dot a}{a})^2)
$$
\begin{equation}
= [d\left(k(d-1)\pm 2(1+\alpha)\right)-A_d]\xi
+[A_d\mp2d(1+\alpha)]\xi_c,
\end{equation}
where we set
\begin{equation}
\frac{\ddot a}{a}=:\pm(1+\alpha)
\end{equation}
respectively for expansion and contraction.
We define the conformal parameter
\begin{equation}
H_c:=\frac{\dot a}{a}=H\,\dot t_s
\end{equation}
given by the Hubble parameter $H$ and the conformal scale factor
$a=e^\gamma=a_1$
to satisfy the differential equation
\begin{equation}
{\dot H_c}+H^2_c=\pm(1+\alpha),
\end{equation}
which
for constant $\alpha$ has
the solutions
\begin{equation}
H_c(t)=\sqrt{1+\alpha}\tanh[\sqrt{1+\alpha}(t-t_0)],
\end{equation}
and
\begin{equation}
H_c(t)=-\sqrt{1+\alpha}\tan[\sqrt{1+\alpha}(t-t_0)].
\end{equation}
respectively to the sign in Eq.
(8.9).

Condition
(8.4) is satisfied, if and only if
\begin{equation}
\xi {>\atop<} \xi_c\frac{A_d\mp 2d(1+\alpha)}{A_d-k d(d\pm k 2(1+\alpha)-1)},
\end{equation}
respectively for
\begin{equation}
(d-3)H^2_c{>\atop<}\pm 2(1+\alpha)+ k (d-1).
\end{equation}

In the following we restrict to the case of constant $\alpha$,
represented by $\alpha=0$. Let us consider the following cases:

Case $k=1,\frac{\ddot a}{a}=-1:$

In this case Eq.
(8.6) reads
\begin{equation}
M^2=d(d-3)\xi
+\frac{1}{2}(d-1)-(\xi-\xi_c)A_d.
\end{equation}
For $d=3$ there is no vacuum instability
(and hence no symmetry breaking) $\forall \xi$,
since $M^2=1$ contradicts the condition
(8.4).

For $d\neq 3$ we find
$$
M^2=d(d-3)[(1-(\frac{\dot a}{a})^2]\xi
+\frac{d-1}{4}[(d-3)(\frac{\dot a}{a})^2+2]
$$
\begin{equation}
=[d(d-3)-A_d]\xi +\xi_c(A_d+2d).
\end{equation}
Therefore condition
(8.4) is satisfied, if and only if
\begin{equation}
\xi {>\atop<} \xi_c\frac{(d-3)(H^2_c-2)}
                        {(d-3)[(H^2_c-1]}
=\xi_c\frac{A_d+2d}{A_d-d(d-3)},
\end{equation}
respectively for $H^2_c{>\atop<}1$ if $d>3$ or
$H^2_c{<\atop>}1$ if $d<3$.

Concerning the latter case especially for $d=2$ we have
$A_2=-2\frac{\dot a^2}{a^2}$
and Eq.
(8.15) becomes
\begin{equation}
M^2=2[(\frac{\dot a}{a})^2-1]\xi
+\frac{1}{4}[2-(\frac{\dot a}{a})^2]
=-(2+A_2)\xi
+\xi_c(4+A_2),
\end{equation}
yielding the condition
\begin{equation}
\xi {<\atop>} \xi_c\frac{H^2_c-2}{H^2_c-1}
=\xi_c\frac{A_2+4}{A_2+2},
\end{equation}
for $H^2_c {>\atop<} 1$ respectively.

Case $k=-1,\frac{\ddot a}{a}=1:$

In this case Eq.
(8.6) reads
\begin{equation}
M^2=-d(d-3)\xi
-\frac{1}{2}(d-1)-(\xi-\xi_c)A_d.
\end{equation}
For $d=3$ it is $M^2=-1$ and the vacuum is unstable
(hence symmetry breaking could occure) $\forall\xi$.

For $d\neq 3$
we find
$$
M^2=-d(d-3)[(1+(\frac{\dot a}{a})^2]\xi
+\frac{d-1}{4}[(d-3)(\frac{\dot a}{a})^2-2]
$$
\begin{equation}
=-[d(d-3)+A_d]\xi +\xi_c(A_d-2d).
\end{equation}
Therefore condition
(8.4) is satisfied, if and only if
\begin{equation}
\xi {>\atop<} \xi_c\frac{(d-3)H^2_c-2}
                        {(d-3)[H^2_c+1]}
=\xi_c\frac{A_d-2d}{A_d+d(d-3)},
\end{equation}
respectively for $d>3$ and $d<3$.

Concerning the latter case especially for $d=2$ we have
$A_2=-2\frac{\dot a^2}{a^2}$
and Eq.
(8.20) becomes
\begin{equation}
M^2=2[(\frac{\dot a}{a})^2+1]\xi
-\frac{1}{4}[2+(\frac{\dot a}{a})^2]
=(A_2+2)\xi
-\xi_c(A_2+4),
\end{equation}
yielding the condition
\begin{equation}
\xi {<} \xi_c\frac{H^2_c+2}{H^2_c+1}
=\xi_c\frac{A_2+4}{A_2+2}.
\end{equation}

\section{ Conclusion}
\setcounter{equation}{0}
We have examined conformally equivalent Lagrangian models with a
scalarfield coupled to geometry.

In Sec. 3 the conformal transformation of the minimally coupling model
(MCM) to the conformal coupling model (CCM) has been performed in
arbitrary dimensions $D$, with the conformal factor and scalar field
in agreement with the results of Ref. \Xa. By Eq. (3.17) the proper
generalization of the scalar field from the conformal coupling case to
that of an arbitrary coupling constant $\xi$ has been found. (Note
that Eq. (5) in Ref. \Pa\ holds only for $\xi=\xi_c=\frac{1}{6}$ with
$D=4$).

In Sec. 4 the geometry has been specialized to multidimensional
cosmological models.

In Sec. 5 we have considered natural time gauges in multi-dimensional
universes: (i) synchronous time, (ii) conformal times of different factor
spaces, (iii) mean conformal time and (iv) harmonic time.

In Sec. 6 the equivalent Lagrangian models of Sec. 3 have been
compared on the level of multidimensional solutions.
For the case of a massless ( $U(\Phi)=0$ ) minimally coupled
scalar field $\Phi$
we found the multidimensional generalization of the classical Kasner solution.
The conformal transformation
of the Kasner solution for the MCM with flat internal spaces $M_i$
yields the nonsingular solution (6.13) for
$\phi^2<\xi^{-1}_c$ and to the singular solution (6.14) for
$\phi^2>\xi^{-1}_c$. This resolution of the scale factor singularity
of the Kasner solution for a proper CCM solution (6.13) confirms
for arbitrary dimension $D$ what has been indicated in Ref. \Ga\ for $D=4$.
At $\phi^2=\xi^{-1}_c$ there is a singularity of the conformal transformation.
The conformal equivalence only holds separately in the ranges
$\phi^2<\xi^{-1}_c$ and $\phi^2>\xi^{-1}_c$.

In the special case of static
internal spaces, we find a minimal scale $a_1(\tau_0)$ at
(harmonic) time $\tau_0$ where the birth of the universe $M$ is happening.
Analytic continuation
of this solution to the Euclidean time region (preserving
real geometry) yields a purely imaginary scalar field.
This solution corresponds to an (instanton) wormhole,
where the scale $a_1(i\tau_0)$ now indicates the throat of the
wormhole.

In Ref. \BlRZ\ these solutions have also been compared to their
quantum counterparts. However,  conformal equivalence transformations
of the classical Lagrangian models and minisuperspace conformal
transformations are conceptually very different proceedures$^{\bRa}$.
This is in analogy to the difference between conformal coordinate
transformations and conformal transformations of Lagrangian models.

In Sec. 7 a number theoretical analysis of the conformal coupling
constant $\xi_c$ indicates some kind of resonance of the coupling for
different dimensions. This  distinguishes the dimensions $3,4,6,10$
and $26$ against all other dimensions $D<82$.

In Sec. 8 it was shown how $\xi_c$ also  enters crucially
in a negative mass condition, necessary
for a vacuum instability. The corresponding
inequalities have been derived for a scalar field
on the background of a multidimensional geometry with static internal spaces.

However further investigations will be required to yield a more  detailed
and more genearal understanding of the conditions for dynamical
or spontaneous compactification.
In Ref. \bBlZ\ it is shown for the model of Eq.
(4.5)
how both types of compactification
can be obtained in the flat case.

\np
\noindent
{\large {\bf Acknowledgements}}
\nl\nl
This work was supported by WIP grant 016659 (U.B.)
and DFG grant Bl 365/1-1 (M.R.).
We are grateful to A. Zhuk for helpful discussions during
our collaboration.
\nl\nl

\bigskip

\noindent
{\large {\bf References}}
\nl\nl
$^{\aBlLP}$  U. Bleyer, D.-E. Liebscher and A. G. Polnarev,
Nuovo Cim. B {\bf 106}, 107 (1991).
(1991).
\nl
$^{\bBlLP}$  U. Bleyer, D.-E. Liebscher and A. G. Polnarev,
{\em Kaluza-Klein Models},
Proc. V$^{th}$ Seminar on Quantum Gravity, Moscow 1990, World Scientific
(1991).
\nl
$^{\Pa}$  D. N. Page, J. Math. Phys. {\bf 32}, 3427 (1991),
\nl
$^{\Sch}$  H.-J. Schmidt, Phys. Lett. B {\bf 214}, 519 (1988).
\nl
$^{\Li}$  J. E. Lidsey, Gen. Rel. Grav. {\bf 25}, 399 (1993),
\nl
$^{\aBlZ}$  U. Bleyer and A. Zhuk, {\em On Scalar Field in Multidimensional
Cosmological Models}, Preprint FUB-HEP/92-23, FU Berlin (1992).
\nl
$^{\Ma}$  K. Maeda, Phys. Rev. D {\bf 39}, 3159 (1989),
\nl
$^{\Xa}$  B. C. Xanthapoulos and Th. E. Dialynas, J. Math. Phys. {\bf 33},
1463 (1992).
\nl
$^{\Ga}$  D. V. Gal'tsov and B. C. Xanthopoulos, J. Math. Phys. {\bf 33},
273 (1992).
\nl
$^{\BlRZ}$  U. Bleyer, M. Rainer and A. Zhuk, {\em Conformal
Transformation of Multidimensional Cosmological Models
and their Solutions}, Preprint FUB-HEP/94-3, FU Berlin (1994).
\nl
$^{\Gi}$  G. W. Gibbons and S. W. Hawking, Phys. Rev. D {\bf 15}, 2752 (1977).
\nl
$^{\Vi}$  A. Vilenkin, Phys. Rev. D {\bf 27}, 2848 (1983).
\nl
$^{\aRa}$  M. Rainer, {\em Projective Geometry for Relativistic Quantum
Physics}, Proc. 23$^{rd}$ Ann. Iranian Math. Conf. (Baktaran, 1992);
J. Math. Phys. {\bf 35}, 646 (1994).
\nl
$^{\Ze}$  H. D. Zeh, {\em The Physical Basis of the Direction of Time},
2nd. ed. Springer-Verlag (Heidelberg, 1991);
\nl
$^{\bRa}$  M. Rainer, {\em Conformal Coupling and Invariance in Arbitrary
Dimensions}, Preprint 94/2, Mathematisches Institut,
Universit\"at Potsdam (1994).
\nl
$^{\bBlZ}$  U. Bleyer and A. Zhuk, {\em Multidimensional Integrable
Cosmological Models with Dynamical and Spontaneous Compactification},
Preprint FUB-HEP/93-19, FU Berlin (1993).
\end{document}